\documentstyle[12pt,epsf,epsfig,wrapfig]{article}  
\begin{document}  
\newtheorem{thm}{Theorem}
\newtheorem{cor}{Corollary}
\newtheorem{Def}{Definition}
\newtheorem{lem}{Lemma}
\begin{center}  
{\large \bf  
Rotational invariance and the spin-statistics theorem.
\ }  
\vspace{5mm}

Paul O'Hara
\\  
\vspace{5mm}  
{\small\it  
Dept. of Mathematics\\ 
Northeastern Illinois University\\ 
5500 North St. Louis Avenue\\
Chicago, Illinois 60625-4699.\\
\vspace{5mm} 
email: pohara@neiu.edu \\}  
\end{center}  
\vspace{10mm}
\begin{abstract}
In this article, the rotational invariance of
entangled quantum states is investigated as a possible cause of the
Pauli exclusion principle. First, it is shown that a certain
class of rotationally invariant
states can only occur in pairs. This is referred to as the
coupling principle. This in turn suggests a natural classification of
quantum systems into those containing coupled states and those that do not.
Surprisingly, it would seem that Fermi-Dirac statistics follows
as a consequence of this coupling while the Bose-Einstein follows by breaking
it. In section 5, the above approach is related to Pauli's original spin-statistics
theorem and finally in the last two sections, a theoretical justification, based on
Clebsch-Gordan coefficients and the experimental evidence respectively,
is presented.\newline 

\end{abstract}

\section {INTRODUCTION}

Rotational invariance in quantum mechanics is usually associated 
with spin-singlet states.
In this article, after having first established a uniqueness theorem relating
rotational invariance and spin-singlet states, a statistical classification
is carried out. In effect, it will be shown that, within the construct of the
proposed mathematical model,
rotationally invariant quantum states can only occur in pairs. These 
pairs will be referred to as isotropically spin-correlated states (ISC) and
will be defined more precisely later on. This in turn will suggest a statistical
classification procedure into systems containing paired states and those that  
do not. It will be shown that a system of
$n$ coupled and indistinguishable states obey the Fermi-Dirac statistic,
while Bose-Einstein statistics will follow when the coupling is broken.

At the same time, it is important to point out that, although, these 
results can be made mathematically compatible
with Pauli's article of 1940 \cite{pauli}, nevertheless
there are differences. In the usual quantum field theory approach, 
Klein-Gordan fields are first quantized and then it is shown that creation operators
commute, as also do annihilation operators. Similarly, when the Pauli principle is
applied to the Dirac field, it is found that the analogous operators anticommute.
However, our approach is different.
We begin, not by focusing on Hamiltonian fields but rather on the rotational
properties of spin. From there we proceed to show that an antisymmetric wave 
function can be associated with rotationally invariant states (singlet states),
while the symmetrical wave function can be associated with the absence of rotationally 
invariant states. Later these rotationally invariant states and non-rotationally
invariant states are related to anticommutative spin operators and commutative
spin operators, respectively.
{\it It follows that our usage of the terms fermions and bosons is not based on
spin value but on rotational properties and correlations between quantum states.}

At first this may seem like a huge break with Pauli's formulation in terms of
anticommutator and commutator relationships and one may even object to giving
new meanings to well established terms. 
However, on a closer analysis
it will be found that mathematically, our results are compatible 
with Pauli's approach,
provided we redefine the spin angular momentum by $S=nL$.
Essentially, we are introducing a scaling factor into the definition of angular
momentum which allows us to compare spin 1/2 particles (represented by n=1)
with spin 1 particles (represented by n=2). Once, this rescaled spin operator 
is introduced, we will find that we are now free to apply the usual quantum field theory
arguments to these rescaled operators. We will find, like Pauli, that 
antisymmetric wave functions can be associated with anticommutative spin operators,
while the symmetric wave function can be associated with commutative operators.
However, because of our rescaled momentum operators, we will no longer be
able to intepret the statistics in terms of 1/2 integer and integer values
but rather in terms of rotationally and non-rotationally invariant states.   

These results can perhaps be best summarized in terms of the following scheme:

Rotationally invariant pairs $\Rightarrow$ antisymmetric wave function
$\Rightarrow$ anticommutator relations.\newline
In contrast, Pauli's approach gives:

1/2 integer spin $\Rightarrow$ antisymmetric wave function
$\Rightarrow$ anticommutator relations.

The parallels are clear but so also are the differences.  For this reason the 
reader should be attentive with the usage of the words fermions and bosons in
this article. Our usage is in full agreement with the conventional usage if we
focus on antisymmetric and symmetric quantum states. It is not the same if we focus
on spin-value.

Finally, we turn to notation. Throughout the paper
$\theta$ will represent a polar
angle lying within a plane such that 
$0\le \theta < 2\pi$. Also denote $|\theta_j-\theta_i|$ by
$\theta_{ij}$ and write $a.e.\ \theta$ for ``$\theta$ almost
everywhere''.\\
$\left|\psi_{1\dots n}(\lambda_1,\dots ,\lambda_n)\right>$ will represent an
n-particle state, where $1\dots n$ represent particles and
$\lambda_1 \dots
\lambda_n$ represent the corresponding states.  However, if there
is no 
ambiguity oftentimes this state will be written in the more compact
form 
$\left|\psi(\lambda_1, \dots ,\lambda_n)\right>\ {\rm\ or\ more\ simply\ as}\ 
\left|\psi\right>.$\\
$s_n(\theta)$   
will represent the spin states of particle $n$ measured in 
direction $\theta$ where $s_n(\theta)=\left|\pm\right>$.
In the case of $\theta =0$,  replace $s_n(0)$ with $s_n$ or by
$\left|+\right>$ or $\left|-\right>$ according to
the context, where $+$ and $-$ represent spin up and spin down
respectively.\\ 
Also let $s^-_n(\theta)$ denote the spin state
{\it orthogonal}  to $s_n(\theta)$.\\ 
The wedge product of $n$ 1-forms is given by: 
\begin{eqnarray} a_1\wedge \dots \wedge a_n=
\frac{1}{n!}\delta^{1\dots n}_{i_1\dots i_r}a^{i_1}\otimes \dots \otimes
a^{i_n}.
\end{eqnarray}
Specifically, 
$a^1\wedge a^2 \wedge a^3= \frac{1}{3!}(a^1\otimes a^2 \otimes
a^3+a^2\otimes a^3 \otimes a^1 + a^3 \otimes a^1 \otimes a^2 - a^2\otimes
a^1\otimes a^3
 -a^1\otimes a^3\otimes a^2 - a^3\otimes a^2 \otimes a^1)=
det(a_1,a_2,a_3)e^1\otimes e^2 \otimes e^3.$

\section{A COUPLING PRINCIPLE}

The concept of isotropically spin-correlated states (to be abbreviated as
ISC) is now introduced. This definition is motivated by the probability 
properties of rotational invariance.
Intuitively, $n$ particles are said to be isotropically spin-correlated, if 
a measurement made in an {\it arbitrary} direction $\theta$ on {\it one}  
of the particles allows us to predict with certainty the spin value of each
of the other $n-1$ particles for the same direction $\theta$.
\begin{Def} Let $H_1\otimes H_2$ be a tensor product of two
2-dimensional inner product spaces. Then $\left|\psi\right>\in H_1\otimes H_2$
is said to be rotationally invariant if
\begin{eqnarray}(R_1(\theta),R_2(\theta))\left|\psi\right>=\left|\psi\right>,
\end{eqnarray} where each 
\begin{displaymath}
R_i(\theta)= 
\left[\begin{array}{cc}
\ \cos(c\theta) & \sin(c\theta)\\
-\sin(c\theta) & \cos(c\theta)
\end{array}\right]
\end{displaymath} 
represents a rotation on the space $H_i$ and $c$ is a constant.\cite{green}
\end{Def}
\begin{Def} Let $H_1, \dots ,H_n$ represent $n$ 
2-dimensional inner product\newline
 spaces. $n$ particles are said to be isotropically 
spin correlated (ISC) if\\
(1) for all $\theta$ the two state $\left|\psi_{ij}\right>\in H_i\otimes H_j$ 
is 
rotationally invariant for all $i,j$ where $i\neq j$ and $1\le i,j \le n$,\\
(2) for all $\theta$ and each $m\le n$ the  state 
$\left|\psi\right>\in H_1\otimes \dots \otimes H_m$
can be written as
\begin{eqnarray}
\left|\psi\right>=\frac{1}{\sqrt2}[s_1(\theta)\otimes s_2(\theta) \dots \otimes
s_m(\theta)\pm
s^{-}_1(\theta)s^{-}_2(\theta)\dots \otimes s^{-}_m(\theta)]
\end{eqnarray}
\end{Def}
Note that it follows from the definition of ISC states that rotationally
invariant states of the form
\begin{eqnarray}
\left|\psi\right>=\frac{1}{2}(\left|+\right>\left|+\right> + \left|-\right>\left|-\right> + \left|+\right>\left|-\right> - \left|-\right>\left|+\right>)
\end{eqnarray}
are excluded. In other words,  the existence of ISC states means that
if we measure the spin state $s_1$ then we have simultaneously measured
the spin state for $s_2 \dots s_n$. It can also be shown by
means of projection operators that the state defined by equation (1) is the
{\it only} state that can be projected onto the state 
$\left|\psi_{ij}\right>\in H_i\otimes H_j$ for each $i, j$. 
This further highlights its significance.

Two examples of ISC states can be immediately given:
\begin{eqnarray} \left|\psi\right>=\frac{1}{\sqrt 2}(\left|+\right>\left|+\right>+\left|-\right>\left|-\right>) \end{eqnarray}
and
\begin{eqnarray} \left|\psi\right>=\frac{1}{\sqrt 2}(\left|+\right>\left|-\right>-\left|-\right>\left|+\right>). \end{eqnarray}
However, what is not apparent is that these are the only ISC states permitted
for a system of n-particles. This is now proven.

First note that in terms of the physics, the existence of ISC states
means that once the spin-state 
of particle 1 is measured along an axis then the values of each of the other 
particle states are
also immediately known along the same axis.

In effect, in order to show that such states exist only for n=2, it is sufficient 
to show that it is impossible to have three such 
particles. This follows, since
the existence of $n$
ISC particles, presupposes the existence of $n-1$ such particles.
Moreover, the proof also throws deeper understanding on
the interpretation of Bell's inequality.\newline
\begin{wrapfigure}{r}{8cm}  
\epsfig{figure=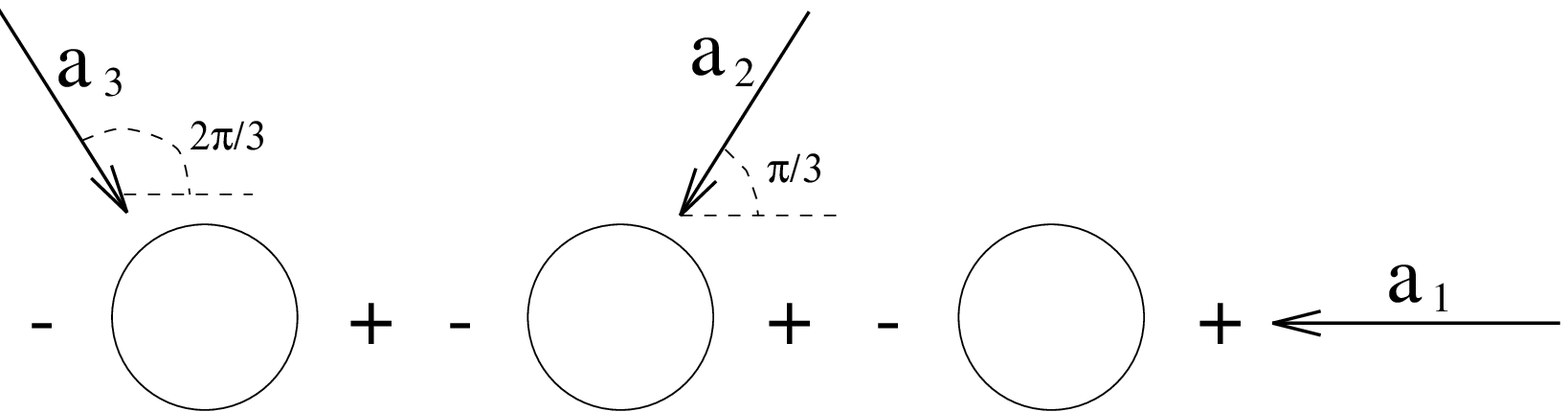,width=8cm}  
{\small Figure 1: Three isotropically spin-correlated
particles.}
\end{wrapfigure}  
In the interest of clarity, assume without loss of
generality, that the three ISC particles are such that
$s_1(\theta_i)=s_2(\theta_i)=s_3(\theta_i)$, for an arbitrary
direction $\theta_i$. This means that the joint spin state for any two of them 
is given by equation (3). 
It follows that three spin measurements can be  
performed, in principle, on the three particle system, in the  
directions  $\theta_i,\ \theta_j, \ \theta_k$. 
Let $(s_1(\theta_i), s_2(\theta_j), s_3(\theta_k))$ represent these 
observed spin values in the three different directions. Recall   
$s_n(\theta)=\pm $ for each n,    
which means that there exists only two possible values for each  
measurement. Hence, for three measurements there are a total of 8 
possibilities in total.  
In particular, following an argument of Wigner \cite{wig}, 
  
\vspace{-6mm}  
  
\begin{eqnarray}  
\;\;\;\;\;\;\{(+,+,-), (+,-,-)\} \subset  
\{(+,+,-),(+,-,-),(-,+,-),(+,-  
,+)\}\end{eqnarray}
implies  
\begin{eqnarray}
P\{(+,+,-),(+,-,-)\} \le  
P\{(+,+,-),(+,-,-),(-,+,-),(+,-,+)\}.  
\end{eqnarray}  
Therefore,  
\begin{eqnarray}\frac 12 \sin^2 \frac {\theta_{ki}}{2}\le \frac 12 \sin^2 \frac
{\theta_{jk}}{2} + \frac 12 \sin^2 \frac {\theta_{ij}}{2},\end{eqnarray}
which is Bell's inequality. Taking
$\theta_{ij}=\theta_{jk}=\frac {\pi}{3}$ and  
$\theta_{ki}=\frac {2\pi}{3}$ gives $\frac 12 \ge \frac
34$, a contradiction. In other words, three  
particles cannot all be in the same spin state with probability  
1, or, to put it another way, isotropically spin-correlated
particles must occur in pairs.\\
\noindent
{\bf Remarks:} (i) If the ISC particles includes the singlet state (4), 
such that 
$(s_1(\theta_i)=+,s_2(\theta_i)= -, s_3(\theta_i)=+)$ (note same
$i$), then regardless of distinguishability,
the spin measurements in the three different
directions $\theta_i$, $\theta_j$, $\theta_k$ can be written as: 
\begin{eqnarray}
\{(+,+,-), (+,-,-)\} \subset  
\{(+,+,-),(+,-,-),(-,-,-),(+,+  
,+)\}.\end{eqnarray}
The previous argument can now be repeated as above.  

(ii) Each of the previous arguments applies also to spin 1
particles, like the photons, provided full angle formulae are
used to derive Bell's inequality, instead of the half-angled formulae. 

(iii) A more rigorous treatment of the above theorem can be found in \cite{had1}.

\section{PAULI EXCLUSION PRINCIPLE}

The above results can be cast into the form of a theorem (already
proven above) which
will be referred to as the ``coupling principle.''
\begin{thm}(The Coupling Principle) Isotropically
spin-correlated particles must occur in PAIRS.\end{thm} 

It follows from the coupling principle that multi-particle systems can be 
divided into two categories, those containing coupled 
particles and
those containing decoupled particles. It now remains
to show that a statistical analysis of these two 
categories, applied to indistinguishable particles, 
generates the Fermi-Dirac and Bose-Einstein statistics respectively.

First note that in the case of ISC particles the two rotationally invariant
states (5) and (6) can be identified with each other by identifying 
a spin measurement of $\pm$ on the second particle in 
the $x$ direction with a spin measurement of $\mp$ in
the $-x$ direction, in such a way as to maintain the rotational invariance. 
In other words, by replacing the second spinor with its spinor conjugate 
\cite{car}, the state  
$\left|\psi\right>=\frac{1}{\sqrt 2}(\left|+\right>\left|+\right>+\left|-\right>\left|-\right>)$
can be written as
$\left|\psi(\pi)\right>=\frac{1}{\sqrt 2}(\left|+\right>\left|-\right>_{\pi}-\left|-\right>\left|+\right>_{\pi})$,
where the $\pi$ subscript refers to the fact that the measurements on
particles 1 and 2 are made in opposite directions, while maintaining the 
rotational invariance \cite{car}.  
The state $\left|\psi(\pi)\right>$ shall be referred to as an
{\it improper} singlet state. Furthemore, without loss of generality
the above identification means
it is sufficient to confine oneselves to {\it singlet} states when 
discussing the properties of ISC particles.
Whether or not improper singlet states (5), actually
exist in nature remains an open question. They can be easily eliminated
by including  the azimuthal angle
in the definition of the rotation operator, in which case only the regular
singlet state would remain. However, we will continue to work with both, for
if such states were actually discovered they would throw new light on the 
``handedness'' problem
and their existence might possibly be linked to parity violation.

It remains to show that the requirement of rotational invariance for ISC
particles generates a Fermi-Dirac type statistic. Before doing so,
it is important to emphasize that the Pauli principle has not been assumed 
but rather is being derived from the  usual form of the principle 
(written in terms of the Slater determinant) by
imposing orbital restrictions on the ISC states. 
The essential ideas are as follows:
the existence of ISC particles means that from the coupling principle (above), 
the wave function can be written uniquely
as a singlet state on the $H_1\otimes H_2$ space. It then follows by
imposing the additional requirement that ISC particles occur only 
in the same orbital, that the usual singlet-state form of the wave
function can be extended to the space ${\cal S}_1\otimes {\cal S}_2$ where
${\cal S}_i=L^2({\cal R}^3)\otimes H_i$. 
Indeed, it would almost seem to be a tautology stemming
from the definition of rotational invariance. 
Nevertheless, the manner in which the notion of ISC states extend to
these spaces needs to be clarified, since
it permits an extension of the results to the usual 
$L^2 \otimes H$ space
associated with quantum mechanics and not just the more restricted spin
spaces. Once this extension is made, proof by induction can be used to derive 
the usual form of the Pauli principle 
associated with the Slater determinant, for an n-particle system.
Moreover, it is also worth noting that the existence of
spin-singlet states in general, and not only in the same orbital, permits 
more general forms of the exclusion principle \cite{had1}.
 
In what remains, let $s_n=s_n(\theta)$ represent the
spin of particle $n$ in the direction $\theta$.  
Also, let $\lambda_n=(q_n, s_n)$ represent the
quantum coordinates of particle $n$, with $s_n$ referring to the
spin coordinate in the direction $\theta$ and $q_n$ representing
all other coordinates. In practice, 
$\lambda_n=(q_n, s_n)$ will represent the coordinates of the particle
in the state $\psi(\lambda_n)$ defined on the Hilbert space
${\cal S}_n= L^2({\cal R}^3)\otimes H_n$,
where $H_n$ represents a two-dimensional spin space of the
particle $n$. This 
distinction will also allow the ket to be written as:
$\left|\psi(\lambda)\right>=\left|\psi(q)\right>s=\left|\psi(q)\right>\otimes s$ 
where $s$ represents the spinor.
With these distinctions made, the notion of orbital is now defined 
and a sufficient condition for obtaining the usual form of the 
Fermi-Dirac statistics within the context of our mathematical model is given. 
\begin{Def}Two particles whose states are given by $\left|\psi(q_1,s_1)\right>$ and\\ 
$\left|\psi(q_2,s_2)\right>$ respectively are said to be in the same q-orbital when
$q_1=q_2$.
\end{Def}
The following lemma allows us to extend the results for ISC particles
defined on the space $H_1\otimes H_2$ to the larger space 
${\cal S}_1\otimes {\cal S}_2$.
As mentioned above,  the Pauli 
principle is not being assumed but rather is being deduced by invoking
rotational invariance of the ISC particles. 
Conversely, if the rotational invariance condition
is relaxed then the Pauli principle need not apply and as a result many
particles can be in the same orbital. 
\begin{lem} Let
\begin{eqnarray}
\left|\psi(\lambda_1,\lambda_2)\right>
&=&c_1\left|\psi_1(\lambda_1)\right> \otimes
\left|\psi_2(\lambda_2)\right>+c_2\left|\psi_1(\lambda_2)\right> \otimes
\left|\psi_2(\lambda_1)\right>\end{eqnarray} 
represent an indistinguishable two particle
system defined on the space ${\cal S}_1\otimes {\cal S}_2$.
If ISC states for a system of two indistinguishable
and non-interacting particles occur only in the same q-orbital  
then the system
of particles can be represented by the Fermi-Dirac statistics.
\end{lem}
{\bf Proof:} The general form of the non-interacting and indistinguishable two
particle state is given by
\begin{eqnarray}
\left|\psi(\lambda_1,\lambda_2)\right>
=c_1\left|\psi_1(\lambda_1)\right> \otimes
\left|\psi_2(\lambda_2)\right>+c_2\left|\psi_1(\lambda_2)\right> \otimes
\left|\psi_2(\lambda_1)\right>\\
=c_1\left|\psi_1(q_1)\right>s_1 \otimes
\left|\psi_2(q_2)\right>s_2+c_2\left|\psi_1(q_2)\right>s_2 \otimes
\left|\psi_2(q_1)\right>s_1\end{eqnarray}
where $c_1,\ c_2$ are constants for all $\lambda_1$ and $\lambda_2$.
Let $q_1=q_2$, then the particles are in the same q-orbital. 
By invoking the ISC condition, it follows from the coupling principle and the 
rotational invariance that
$c_1=-c_2$. Therefore,
\begin{eqnarray}\left|\psi(\lambda_1,\lambda_2)\right>=\frac{1}{\sqrt 2}[\left|\psi_1(\lambda_1)\right> \otimes
\left|\psi_2(\lambda_2)\right>-\left|\psi_1(\lambda_2)\right> \otimes
\left|\psi_2(\lambda_1)\right>]\end{eqnarray}
by normalizing the wave function. 
The result follows. QED\\
\noindent
{\bf Remarks}: (i)Singlet states composed of particles of spin n can also be 
handled by the above theory. For example, the singlet state 
\begin{eqnarray}\left|0,0\right>=\frac{1}{\sqrt 5}\left(\left|2,-2\right>-\left|1,-1\right>+\left|0,0\right>
-\left|-1,1\right>+\left|-2,2\right>\right)\end{eqnarray}
can be decomposed into two irreducible ISC representations of the form
$\left|2,-2\right>-\left|-2,2\right>$ and 
$\left|1,-1\right>-\left|-1,1\right>$ and the above theory can then be 
applied to these states. The remaining $\left|0,0\right>$ state can be treated
as a bosonic state (see Section 4).\\
(ii) The above lemma also applies to improper singlet states,
in other words to particles whose spin correlations are given by equation (5). 
This can be done by correlating a measurement in
direction $\theta$ on one particle, with a measurement in
direction $\theta + \pi$ on the other.  In this case, the state
vector for the parallel and anti-parallel measurements will be
found to be: 
\begin{eqnarray}\left|\psi(\lambda_1,\lambda_2)\right>=\frac{1}{\sqrt
2}[\left|\psi(\lambda_1)\right>
\otimes
\left|\psi(\lambda_2(\pi))\right>-\left|\psi(\lambda_2)\right> \otimes
\left|\psi(\lambda_1(\pi))\right>],\end{eqnarray}
where the $\pi$ expression in the above arguments refers to the
fact that the measurement on particle two is made in the opposite
sense to that of particle one.\\
(iii)If the coupling condition (rotational invariance) is removed then a 
Bose-Einstein
statistic follows and is of the form:
\begin{eqnarray}\left|\psi(\lambda_1,\lambda_2)\right>=\frac{1}{\sqrt
2}[\left|\psi(\lambda_1)\right>
\otimes
\left|\psi(\lambda_2)\right>+\left|\psi(\lambda_2)\right>\otimes
\left|\psi(\lambda_1)\right>].\end{eqnarray}
This will be discussed in more detail later. See, for example, Corollary 1
following Theorem 3. 

Our next theorem gives the usual formulation of the Pauli exclusion principle
for exchangeable particles.
\begin{thm} (The Pauli Exclusion Principle) A
sufficient condition for a state, representing n-cyclically permutable and
non-interacting particles, defined on the space ${\cal S}_1\otimes \dots
{\cal S}_n$
to exhibit Fermi-Dirac statistics is that it contain spin-coupled
q-orbitals.\end{thm}
{\bf Remark:} A system of n-cyclically permutable particles will be referred to
as n indistinguisable particles.\newline
\noindent
{\bf Proof:} It is sufficient to work  with three particles, but it should be
clear that the argument can be extended by induction to an n-particle
system.  Consider a system of three
indistinguishable particles, containing spin-coupled particles.
Using the above notation and applying equation (14) of
Lemma 1 to the coupled particles in the second line below, gives:
\begin{eqnarray*}\left|\psi (\lambda_1, \lambda_2,
\lambda_3)\right>&=&
\frac{1}{\sqrt 3}
\{\left|\psi(\lambda_1)\right>\otimes \left|\psi(\lambda_2,\lambda_3)\right> +
\left|\psi(\lambda_2)\right>\otimes \left|\psi(\lambda_3,\lambda_1)\right>\ \\
&\ &\hspace{5mm}
+\left|\psi(\lambda_3)\right>\otimes \left|\psi(\lambda_1,\lambda_2)\right>\}\\
&=&\frac{1}{\sqrt {3!}}\{\left|\psi(\lambda_1)\right>\otimes
[\left|\psi(\lambda_2)\right>\otimes \left|\psi(\lambda_3)\right>-
\left|\psi(\lambda_3)\right>\otimes \left|\psi(\lambda_2)\right>]\\ 
&\ & +\left|\psi(\lambda_2)\right>\otimes [\left|\psi(\lambda_3)\right>\otimes
\left|\psi(\lambda_1)\right>-
\left|\psi(\lambda_1)\right>\otimes \left|\psi(\lambda_3)\right>]\\ 
&\ &+\left|\psi(\lambda_3)\right>\otimes [\left|\psi(\lambda_1)\right>\otimes
\left|\psi(\lambda_2)\right>-
\left|\psi(\lambda_2)\right>\otimes \left|\psi(\lambda_1)\right>]\}\\
&=&\sqrt{3!}\left|\psi_1(\lambda_1)\right>\wedge \left|\psi_2(\lambda_2)\right>\wedge
\left|\psi_3(\lambda_3)\right>,
\end{eqnarray*} where $\wedge $ represents the wedge product.
Thus the wave function for the three indistinguishable particles
obeys Fermi-Dirac statistics.  The n-particle case follows by
induction. QED\\

For example, consider
the case of an ensemble of 2n identical non-interacting particles
with discrete energy levels
$E_1, E_2, \dots $, satisfying the Fermi-Dirac statistics as above then 
all occurances
of such
a gas would necessarily have a twofold degeneracy in each of the discrete 
energy levels and the lowest energy would be given by
\begin{eqnarray}E=2E_1+2E_2+2E_3+\dots +2E_n.\end{eqnarray}  

To conclude this section, note Lemma 1 and Theorem 2 express a Pauli type  
exclusion principle which follows naturally from the
coupling principle, or equivalently the rotational invariance. In the next
section it  will be
shown that if the rotational invariance is removed and replaced with the 
requirement
that indistinguisable states be equally likely, then the system will obey a 
Bose-Einstein statistic.  

\section{BOSE-EINSTEIN STATISTICS}

In the above discussion rotational invariance has played a key role
in formulating a Fermi-Dirac 
statistic for multi-particle ISC systems as defined by definitions 1, 2 and
3. Indeed,  from the perspective of this paper, it would seem to be the 
underlying cause of
the Pauli exclusion principle.  It now remains to investigate the statistics of 
multiparticle systems when this condition is relaxed.  As noted previously
the rotational invariance implies that ISC particles can be written in the
form of a singlet state, either proper or improper.  Moreover, the definition
of indistinguishability (cf remark to Theorem 2) means that a uniform probability
law is assigned to each orthonormal basis. Physically, this means that there is no  
bias in favor of any of the
components of the permutable states. For example, if
\begin{eqnarray}\left|\psi(\lambda_1,\lambda_2)\right>=a\left|\lambda_1\right>\otimes \left|\lambda_2\right>+b\left|\lambda_2\right>
\otimes \left|\lambda_1\right>\end{eqnarray}
is permutable, then $|a|^2=|b|^2$; otherwise if $|a|>|b|$ (respectively  $|b|>|a|$) 
there 
would be a bias in favor of the state associated with $a$ (respectively $b$)
which, together with the law of large numbers,  could then be used to 
partially distinguish the states.
\begin{thm} Permutable states for a system of $n$ non-interacting particles,
defined on the space ${\cal S}_1\otimes \dots \otimes{\cal S}_n$, obey either 
the Fermi-Dirac or the
Bose-Einstein statistic.
\end{thm}
{\bf Proof:} Let 
\begin{eqnarray}\sigma\left|\psi(\lambda_1, \dots ,\lambda_n)\right>=\sum^{n!}_{i=1}c_i
\left|\psi(\sigma_i \lambda_1)\right>\otimes \dots 
\otimes \left|\psi(\sigma_i \lambda_n)\right>\end{eqnarray}
where $\sigma_i$ represents a permutation of the particles in the states
$\lambda_1, \dots , \lambda_n$. We now claim that if the system of 
indistinguishable particles are not in the Fermi-Dirac state then
\begin{eqnarray}\sigma(\left|\psi(\lambda_1, \dots ,\lambda_n)\right>=\frac{1}{\sqrt n!}\sum^{n!}_{i=1}
\left|\psi(\sigma_i \lambda_1\right>\otimes \dots 
\otimes\left|\psi(\sigma_i\lambda_n)\right>.\end{eqnarray}
First note that if $c_i=\frac{1}{\sqrt n!}$ and 
$c_{i+1}=-\frac{1}{\sqrt n!}$
for each $i$, then Fermi-Dirac statistics results. Hence, assume that there
is {\it not} an exact pairing and that $c_1=\dots c_i=\frac{1}{\sqrt n!}$ 
where either
$i>\frac{n!}{2}$ or $i< \frac{n!}{2}$
and $c_{i+1}=\dots =c_{n!}=-\frac{1}{\sqrt n!}$.
Then taking $\lambda_1=\lambda_2$, many of the terms on the right hand 
side (in fact $\min\{2i, 2(n!-i)\}$ terms) will cancel, leaving only the excess unpaired
positive (negative) terms.  If the remaining number of terms in the
expansion is less than $n!$ then $\left|\psi(\lambda_1,\dots,\lambda_n)\right>$ is
{\it not} invariant under the complete set of permutations, which is a 
contradiction. It follows that the
number of terms must be $n!$ and nothing vanishes. Hence
$\left|\psi(\lambda_1,\dots ,\lambda_n)\right>$ exhibits Bose-Einstein statistics.
The result follows. QED

It might be instructive to apply the above theorem to a three particle wave
function that is {\it not} of the above type. Consider:
\begin{eqnarray*}\left|\psi (\lambda_1, \lambda_2,
\lambda_3)\right>
&=&\frac{1}{\sqrt {3!}}\{\left|\psi(\lambda_1)\right>\otimes
[\left|\psi(\lambda_2)\right>\otimes \left|\psi(\lambda_3)\right>+
\left|\psi(\lambda_3)\right>\otimes \left|\psi(\lambda_2)\right>]\\ 
\ & &+\left|\psi(\lambda_2)\right>\otimes [\left|\psi(\lambda_3)\right>\otimes
\left|\psi(\lambda_1)\right>+
\left|\psi(\lambda_1)\right>\otimes \left|\psi(\lambda_3)\right>]\\ 
\ &&+\left|\psi(\lambda_3)\right>\otimes [\left|\psi(\lambda_1)\right>\otimes
\left|\psi(\lambda_2)\right>-
\left|\psi(\lambda_2)\right>\otimes \left|\psi(\lambda_1)\right>]\}.
\end{eqnarray*} 
On putting $\lambda_1=\lambda_2$,
\begin{eqnarray*}\left|\psi (\lambda_1, \lambda_2,
\lambda_3)\right>
&=&\frac{1}{\sqrt {3!}}\{\left|\psi(\lambda_1)\right>\otimes
[\left|\psi(\lambda_2)\right>\otimes \left|\psi(\lambda_3)\right>+
\left|\psi(\lambda_3)\right>\otimes \left|\psi(\lambda_2)\right>]\\ 
\ & &+\left|\psi(\lambda_2)\right>\otimes [\left|\psi(\lambda_3)\right>\otimes
\left|\psi(\lambda_1)\right>+
\left|\psi(\lambda_1)\right>\otimes \left|\psi(\lambda_3)\right>] 
\end{eqnarray*}
which is not invariant under permutations.
\begin{cor} Let
\begin{eqnarray}\left|\psi(\lambda_1,\dots \lambda_n)\right>
&=&\sum^{n!}_1 c_i\left|\psi(\sigma \lambda_1)\right> \otimes \dots
\left|\psi(\sigma \lambda_n)\right>,\end{eqnarray}
where $c_i$ are independent of
$\lambda_i$, represent an indistinguishable $n$-particle
system defined on the space ${\cal S}_1\otimes \dots \otimes {\cal S}_n$,
such that no two states are ISC, then
this  system
of particles can be represented by the Bose-Einstein statistics.
\end{cor}
{\bf Proof:} Normalizing the wave function and using indistinguishability
gives $c^2_i=c^2_j$, for each $i$ and $j$.  If $c_i=-c_{i+1}$
then each q-orbital
would be a spin-singlet
state.  But this is not so.  Hence $c_i=c_j$ by the previous lemma 
and the result follows. QED
\vskip 7pt 
Denote the set of permutations that leave invariant the Bose-Einstein and 
Fermi-Dirac
statistics by $s_n$ and $a_n$, respectively. It follows that certain types
of mixed statistics can be now described.  For example, the 2-electrons
of a helium atom considered together with the 3-electrons in a lithium 
atom obey $a_2\otimes a_3$ statistics, while the 5 electrons in the boron atom
obey $a_5$ Fermi-Dirac statistics.  The electrons in three
different helium atoms obey $a_2\otimes a_2 \otimes a_2$ if the helium atoms
are considered distinguishable and $s_3\circ (a_2\otimes a_2\otimes a_2)$
if the atoms are indistinguishable.
Finally, if we consider collectively the $n$ distinct electrons in $n$
indistiguishable hydrogen atoms, then these $n$ electrons can be
described with $s_n$ Bose-Einstein statistics.

\section {A CONTRAST WITH PAULI'S\\APPROACH}

It now remains to discuss the above mathematical results from the perspective
of Pauli's famous paper on spin-statistics \cite{pauli} and in the overall 
context of the experimental evidence (discussed next section). 

In Pauli's paper the distinction between the two statistics is made
by distinguishing between operators obeying commutator rules and those
obeying anticommutator rules. Moreover, by defining the Lie algebra for
spin by $[L_i,L_j]=i\epsilon_{ijk}L_k$ fermions are identified with
particles of half-integral spin. Microcausality then guarantees that particles
of integral spin cannot be quantized as fermions and vice-versa. 
In contrast, the model presented in this paper suggests that rotational 
invariance, and not spin value, 
underlies Fermi-Dirac statistics. 

Can the two points of view be reconciled?
The answer is yes, provided we agree to slightly modify the Lie algebra by 
introducing a scaling parameter.  Specifically, re-define the spin angular momentum 
operator by $S_i=nL_i$, where n is an integer, and define the subsequent Lie Algebra
by $[S_i,S_j]=in\epsilon_{ijk}S_k$. Note immediately that the inclusion of the integer
$n$ permits particles of integer spin as well as particles of 1/2-integer spin to be
subjected to anticommutator relations and to obey Fermi-Dirac statistics.
Specifically, when n=1 we obtain the
usual relationship for spin 1/2 particles, whereas for n=2 we obtain the usual
properties for a spin 1 photon and for n=4, the properties of the spin 2 gravitons.

Now, consider a two particle system $\left|\psi(\lambda_1)\right>\otimes 
\left|\psi(\lambda_2)\right>\in {\cal S}_1\otimes{\cal S}_2$ where
${\cal S}_1={\cal L}^2({\cal R}^3)\otimes H_1$ and
${\cal S}_2={\cal L}^2({\cal R}^3)\otimes H_2$ respectively. 
Note that each ket $\left|\psi(\lambda)\right> \in \cal S$ can be written as
$\left|\psi(q_1)\right>\otimes s$, where $s$ is a spinor (page 8). Also, let 
$\vec S_1=(S_i(q_1),S_j(q_1),S_k(q_1))$ and $\vec S_2=(S_i(q_2),S_j(q_2),S_k(q_2))$ 
be spin operators defined on the Hilbert
spaces $H_1$ and $H_2$ respectively. 
We have already seen (Cor. 1) that Bose-Einstein statistics follow when NO two
states are ISC. If we further assume that non-ISC states are statistically independent of
each other, then it would seem  that their corresponding spin operators are best
represented by the operators $S_i(q_1)\otimes I_2$
and $I_1\otimes S_i(q_2)$, where $I_i$ represents an identity operator.
It follows, trivially, that $[S_i(q_1)\otimes I_2,I_1\otimes S_j(q_2)]=0$, which
means that in the case of Bose-Einstein statistics, spin operators must commute.
On the other
hand, in contrast to the triplet state, the singlet state defines a rotationally
invariant state and obeys a Fermi-Dirac statistic. Once again, let 
$\left|\psi(\lambda_1,\lambda_2\right>\in {\cal S}_1\otimes {\cal S}_2$
represent the spin-singlet state of two particles. Note that 
the perfect correlations between them allows us to
put $H_1=H_2=H$ and to identify $\vec S_1$ and $\vec S_2$ as follows: 
Let $s_1(\theta)$ and $s_2(\theta)$ represent the spin states for
particles 1 and 2 respectively, then for an arbitrary angle $\theta$ there exists
a unit vector $\vec n(\theta)$ such that $\vec S_1.\vec n(\theta)(s_1(\theta))
=\pm s_1(\theta)$ if and only if
$\vec S_2.\vec n(\theta)(s_2(\theta))=\mp s_2(\theta)$. This relationship allows us 
to identify $s_2(\theta)$ with the orthogonal complement $s^-_1(\theta)$ of
$s_1(\theta)$ and to put  $\vec S_1=\vec S_2$.
Hence,
\begin{eqnarray*} 
[S_i(q_1),S_j(q_2)]s 
&=& [S_i(q_1),S_j(q_1)]s\\
&=& in\epsilon_{ijk}S_k(q_1)s. 
\end{eqnarray*}
Consequentally, Fermi-Dirac statistics imply spin operators must anticommute
and non-local events in the form 
of spin-singlet states need to be quantized according to the anticommutator rule.
Moreover, since\newline  
$S_i(q_1)S_j(q_2)\neq 0$ and $S_i(q_1)S_j(q_2)+S_j(q_2)S_i(q_1)=0$,
and the above identification is only valid for singlet states, it follows that
bosons can never be fermions and fermions can never be bosons. Also as a special case,
fields quantized by the anticommutator 
rule {\it cannot} be quantized with commutators.

In conclusion, note that Bose-Einstein statistics not only presuppose local and 
independent probability events, but also commuting spin operators. In contrast,
Fermi-Dirac statistics presuppose singlet-states which define non-local interactions
by way of perfect correlations and also imply anticommutating spin operators.
Moreover, since bosons can never be fermions and vice-versa, 
it follows from the above that particles cannot be
simultaneously in non-local and local states, cannot be simultaneously subjected to
the rules of conditional and independent probability, cannot be simultaneously
rotationally invariant and non-invariant. However, once a measurement is performed on the
singlet state then the perfect correlation can be broken, and as a result 
the Fermi-Dirac state can be changed into a Bose-Einstein state; the
singlet state can become two independent states. Finally, note that spin value is no 
longer the essential characteristic of the spin-statistics theorem.

\section{CLEBSCH-GORDAN COEFFICIENTS}

In this section we calculate Clebsch-Gordan coefficients for pairs of photons
and pairs of deuterons to further justify  
the spin-statistics theorem, as presented in this paper.

Consider two photons. Let $s=s_1+s_2$ represent their joint spin values, 
and denote
their joint state by $\left|llsm\right>$. Then
three possible states emerge for $s=2$:
\begin{eqnarray}
\left|2,2\right>&=&\left|1,1\right>\\
\left|2,0\right>&=&\left<1,-1|2,0\right>\left|1,-1\right>+
\left<-1,1|2,0\right>\left|-1,1\right>\\
\left|2,-2\right>&=&\left|-1,-1\right>,
\end{eqnarray}
which defines the triplet state. Also for $s=0$ we obtain
\begin{eqnarray}
\left|0,0\right>&=&\left<1,-1|0,0\right>\left|1,-1\right>+
\left<-1,1|0,0\right>\left|-1,1\right>,
\end{eqnarray}
which defines the singlet state.
Note that the state represented by equation (24)
can be calculated directly by means of C-G coefficients, by recalling that
$S_i=nL_i$ and by observing that for $n=2$
\begin{eqnarray}\left<s,m|S^{\mp}S^{\pm}|s,m\right>=
4\left<s,m|L^{\mp}L^{\pm}|s,m\right>,\end{eqnarray}
from which it follows that
\begin{eqnarray}S^{\pm}\left|s,m\right>=2\hbar[(s\mp m)(s\pm m+1)]^{1/2}
\left|s,m\pm 2\right>.\end{eqnarray}
In particular, if we set  $S^-=S^-_1+S^-_2$ then
\begin{eqnarray}2L^-\left|2,2\right>=
2\hbar[(2+2)(2-2+1)]^{1/2}\left|2,0\right>\end{eqnarray}
and 
\begin{eqnarray}
\left|2,0\right>=\frac{1}{\sqrt2}\left|-1,1\right>+
\frac{1}{\sqrt2}\left|1,-1\right>.\end{eqnarray}
Also if we assume orthogonality of the different states then 
$\left<2,0|0,0\right>=0$ implies 
\begin{eqnarray}
\left|0,0\right>&=&\frac{1}{\sqrt 2}\left|1,-1\right>-
\frac{1}{\sqrt 2}\left|-1,1\right>.
\end{eqnarray}

The deuteron is likewise a spin-1 particle. However in this case, the spin 0 
case can be observed. Moreover, a calculation of the C-G coefficients for a
pair of deuterons says a lot about the probability
weightings associated with the $\left|1\right>$, $\left|0\right>$,
$\left|-1\right>$ states of an individual deuteron. Direct calculation gives:
\begin{eqnarray}
\left|2,2\right>&=&\left|1,1\right>\\
\left|2,1\right>&=&\frac{1}{\sqrt 2}\left|1,0\right>+
\frac{1}{\sqrt 2}\left|0,1\right>\\
\left|2,0\right>&=&\sqrt{\frac{2}{3}}\left|0,0\right>+\frac{1}{\sqrt 6}
\left|1,-1\right>+\frac{1}{\sqrt 6}\left|-1,1\right>\\
\left|2,-1\right>&=&\frac{1}{\sqrt 2}\left|-1,0\right>+
\frac{1}{\sqrt 2}\left|0,-1\right>\\
\left|2,-2\right>&=&\left|-1,-1\right>.
\end{eqnarray}
On the other hand, the probabilities associated with the C-G coefficients
of the states can be calculated directly from conditional probability theory,
\footnote{Recall that for two events $A$ and $B$ defined on a finite sample 
space $S$, the conditional probability of $A$ given $B$ is denoted by $P(A|B)$
and $P(A|B)=P(A\cap B)/P(B)$ provided $P(B)\neq 0$.}
provided the spectral distribution of an individual deuteron
has a probability distribution of 1/4, 1/2, 1/4 and not 1/3, 1/3, 1/3, which
is the current belief.

In particular, let $M_i$ where $i=1,2$ be a random variable associated with 
the spin of two independent deuterons such that
\begin{eqnarray}P(M_i=1)=P(M_i=0)=P(M_i=-1)=\frac{1}{3}.\end{eqnarray}
Let $M=M_1+M_2$ be the sum of the spins. Note that $M$ too is a random variable
with values $2, 1, 0, -1, -2$. 
Then the conditional distribution
for the state $\left|2,0\right>$ associated with the two independent deuterons 
gives 
\begin{eqnarray} P(M_1=0,M_2=0|M=0)&=&P(M_1=1,M_2=-1|M=0)\\
&=&P(M_1=-1,M_2=1|M=0)=\frac{1}{3}.\end{eqnarray}
However, this distribution is clearly different from the C-G calculations 
which gives 
$(\left<0,0|2,0\right>)^2=\frac{2}{3}$,
$(\left<1,-1|2,0\right>)^2=\frac{1}{6}$, 
$(\left<-1,1|2,0\right>)^2=\frac{1}{6}$. On the other hand, if  
\begin{eqnarray}P(M_i=1)=P(M_i=-1)=\frac{1}{4}\ \ \ 
P(M_i=0)=\frac{1}{2}\end{eqnarray}
as suggested by Theorem 2, then direct calculation using conditional probability,
shows that
\begin{eqnarray}&&P(M_1=0,M_2=0|M=0)=\frac{2}{3}\end{eqnarray} and
\begin{eqnarray}P(M_1=1,M_2=-1|M=0)=P(M_1=-1,M_2=1|M=0)=
\frac{1}{6},\end{eqnarray}
which coincides with the C-G calculation.

\section{EXPERIMENTAL EVIDENCE}

The experimental justification for accepting the new form of the
spin-statistics theorem would appear to come from a wide range of physical
phenomena. First, note that the existence of photons in the spin-singlet
state seems to support the above formulation.
Secondly, we will argue that the new approach offers a more unified and 
coherent explanation of the phenomenon of paramagnetism. Thirdly, the existence 
of Cooper pairs in superconductivity can be explained as a specific instance of
ISC particles. Fourthly, we discuss baryonic
structure from the new perspective. Finally, in keeping with
the tradition of theoretical physics, a  prediction will be made about the
probability distribution for the spin decomposition of a beam of ionized
deuterons,  a prediction which will distinguish it from the current theory.

(1)  Rotational invariance demands the wave function for spin-singlet-state 
photons to be of the form
$\left|\psi\right> = \frac{1}{\sqrt 2}(\left|+\right>\left|-\right>-
\left|-\right>\left|+\right>).$ Since this is an antisymmetric state then
by definition it obeys
a Fermi-Dirac type statistic for the Hilbert space under discussion. 
Moreover, it has 
already been pointed out in Section 5
that singlet states, and not spin value, 
are essential to defining Fermi-Dirac statistics.
Spin-singlet-state photons were at the heart of 
the  Aspect experiment \cite{asp} 
and hence their existence has already been verified.
The exclusion principle is then a tautology in the sense that while
photons are in a spin-singlet state then both of them cannot be in the
same state.
Note, however, that the fermionic state of photons can easily be destroyed 
by experiment and forced into a Bose-Einstein state. It is a trite (but
nevertheless valid) application of the exclusion principle, as formulated in this paper.

(2) The theory of paramagnetism yields two different equations
for the magnetic susceptibility, one given by the classical Langevin (Curie)
function which makes no reference to the Pauli exclusion principle and the
other which is derived as a direct application of the exclusion 
principle. It would appear that our formulation of the exclusion principle
gives an equally apt understanding of the phenomenon and would 
appear to further clarify Pauli's explanation, by focusing on the unique
role of the non-spin-singlet states. 
Specifically, when the magnetic field is turned on, 
the spin up component of the spin-singlet state has its energy 
shifted down by $\mu B$ while the spin down component has its energy
shifted up by $\mu B$ with the spins being aligned 
into  parallel and anti-parallel states, resulting in a net contribution to
the magnetic field of 0. 
Hence, the paired elecrons contribute nothing to the magnetic 
susceptibility.  The remaining unpaired electrons act in such a way that 
there is an excess of electrons in the spin up state over the spin down
state, in order to maintain the common electrochemical potential. Specifically, 
if we  let $g(\epsilon)$ be the electron density of
available states per unit energy range then the total excess energy is given
by $g(\epsilon_F)\mu B$, provided $\mu B<<\epsilon_F$, which is
Pauli's result for paramagnetism.  It should also be pointed out that 
from the perspective of Pauli's version of the spin-statistics theorem, 
half-integral-spin particles such as electrons  or 
gaseous-nitric-oxide (NO) molecules remain as fermions regardless of  
thermodynamic considerations or of the state they occupy.
However, it is generally taken for granted 
\cite{hall} that as $kT>>\epsilon_f$ the Pauli principle no longer applies 
and the magnetic susceptibility is in this case best estimated by using the
Boltzmann statistics.  From the perspective of this article, 
this gives rise to the ambivelant situation of referring
to particles as fermions, although they are no longer subjected to the Pauli  
exclusion principle.  With our approach, this ambiguity is removed and  
a more natural and coherent explanation of the transition from Fermi-Dirac to
Boltzmann statistics is forthcoming. Essentially,
the Boltzmann statistics emerge when the spin coupling which seems to be the 
underlying cause of the Fermi-Dirac statistics is first broken
and the particles  then move apart to become distinguishable and 
statistically independent.
This breaking of the coupling occurs naturally when the 
temperature is raised, and they become distinguishable and independent 
when the distance separating the particles becomes large enough to overcome
interactions between the particles. As a result, the particles obey Boltzmann
statistics and 
``the Curie law applies to paramagnetic atoms in a low density gas, just
as to well separated ions in a solid..."\cite{hall}.

Admittedly, other viewpoints may be adequate and someone may believe that 
fermions remain
so, even when the statistical constraints have been removed. However, the purpose 
of the above description is to show that an alternative consistent position, compatible with
the claims of the paper, can be given. 

(3) The existence of Cooper pairs as spin-singlet states in the theory of
superconductors is another instance of the coupling principle at work. 
Moreover, the fact that $2n$ superconducting electrons exhibit the statistics 
of $n$ boson pairs and $\it not$ the usual $a_{2n}$ Fermi-Dirac statistics
\cite{hall}, normally associated with the exclusion principle, 
again suggests that the
current definition of bosons and fermions in terms of quantum number is 
inadequate. In contrast, this paper classifies particles into coupled or
decoupled particles and then permits various statistics to emerge in
accordance with the degree of indistinguishability that is imposed on the 
system. When complete indistinguishability is imposed on the system, then
Fermi-Dirac or Bose-Einstein statistics will ensue according as to whether
the system permits coupled (Theorem 2) or only decoupled particles (Cor 1),
respectively. On the other hand, if complete indistinguishability is relaxed in
favor of some type of partial indistinguishability (as with Cooper pairs),
we obtain different types of mixed statistics. For example, if we denote by $s_n$
and $a_n$ the set
of permutations that leave invariant the Bose-Einstein and Fermi-Dirac statistics,
then as previously pointed out in Section 4, the 2-electrons in a helium atom taken
together with the 3-electrons of a lithium atom obey $a_2\otimes a_3$ statistics,
in contrast to the 5 electrons in the boron atom that obey $a_5$ Fermi-Dirac statistics.
Also, if we consider collectively the $n$ distinct electrons in $n$ indistinguishable
hydrogen atoms, then these $n$ electrons can be described with $s_n$ Bose-Einstein
statistics, since there is no electron pairing.

(4) Spin $\frac 32$ baryons may be viewed as
excited states of spin $\frac 12$ baryons.  In particular, from the perspective
of the new approach, it is impossible by Theorem 1 that a spin $\frac 32$
baryon be composed of three ISC quarks. It could be argued that it contains a
pair of ISC quarks in an improper singlet state (page 7), with the third quark
being uncorrelated with these two. However, unless there is experimental evidence
to suggest otherwise, this seems unlikely. The other alternative is to view a spin 
$\frac 32$ baryon
as composed of three quarks with uncorrelated spin
states (statistically independent), and to view the spin $\frac 12$ baryon as composed
of a pair of quarks in a singlet state. Moreover, the need of color to explain
the structure of $\Delta^{++}$ and $\Omega^-$ particles now becomes both  
unnecessary and inadequate\cite{bar}. The coupling principle forbids
the three quarks composing both the $\Delta^{++}$ and $\Omega^-$ particles to
exist as ISC particles, but rather suggests that they are statistically independent.  
Color fails to fully address this issue.
Of course, this does not preclude the use
of color to give ``colorless'' baryons. \cite{qui}
   
(5) It is well known that the deuteron ion is in a spin-triplet
state. Denote the possible observed spin values $X$ by +1, 0, -1 respectively.  
Conventional quantum mechanics
predicts that $P(X=+1)=P(X=0)=P(X=-1)=\frac 13$. On the other hand, if we
assume that the absence of the spin-singlet state for deuteron ions means
that the Bose-Einstein triplet state is composed of two independently 
distributed spin $\frac 12$ particles then the model proposed in this paper
predicts $P(X=+1)=P(X=-1)=\frac 14$ and
$P(X=0)=\frac 12$, using an argument based on Clebsch-Gordan coefficients
\cite{cg}. This should be testable by passing a beam of neutral 
deuteron atoms (not molecules) through a Stern-Garlach apparatus.
\vskip 5mm
\noindent
{\bf Remark:} Strictly speaking, the failure of particles to form a spin-singlet 
state does not necessarily mean that the subsequent spin values of the triplet 
state are governed by the laws of independent probability. It may mean that
there is some type of dependent but non-deterministic relationship between 
the particles.
This further highlights the importance of performing an experiment 
like that described above. If decoupled spin states imply statistical
independence, then classification procedures become very simple. On the other
hand, if statistical independence fails to be observed then
the Bose-Einstein type statistic would have to be further sub-classified.

\section {CONCLUSION} 

In this paper a ``spin-coupling principle'' is derived which suggests a 
statistical classification of particles in terms of ISC states (spin-entangled
pairs) and non ISC states.
These ISC states appear to unify our understanding of atomic
orbitals, covalent bonding, paramagnetism, superconductivity, baryonic 
structure and so on. In summary, subatomic particles seem to form 
entangled {\it pairs} whenever they are free to do so and there appears to be a 
universal principle at work, although the mechanism behind this coupling
would need to be investigated further.

Secondly, in contrast to the current paper, Pauli's version of the 
spin-statistics theorem imposes many other conditions on his particle system
including Lorentz
invariance, locality ( ``measurements at two space points with a space-like 
distance can never disturb each other''\cite{pauli}, charge and 
energy densities.
However, the imposition of such extra conditions would seem to 
be unnecessary in the light of our current understanding of entanglement.
 
Finally, note that a connection between Bell's inequality\cite{bell}
and rotational invariance has been established.

\vspace{3mm}

\end{document}